\documentclass[reprint,amsmath,amssymb,aps]{revtex4-1}

\usepackage{graphicx}% Include figure files
\usepackage{dcolumn}% Align table columns on decimal point
\usepackage{bm}% bold math
\usepackage{amsmath,amssymb,graphicx}
\usepackage{graphicx}
\usepackage{bm}
\usepackage{float,placeins}
\usepackage{adjustbox}
\usepackage{mathptmx}
\usepackage{epstopdf}
\usepackage{booktabs}
\usepackage{multirow}
\usepackage{hhline}
\usepackage{verbatim}
\usepackage{subfigure}
\usepackage{soul}
\usepackage{blindtext}
\usepackage[colorlinks=true,linkcolor=blue]{hyperref}
\hypersetup{
    colorlinks=true,    % false: boxed links; true: colored links
    linkcolor=blue,       % color of internal links
    citecolor=blue,       % color of links to bibliography
    filecolor=red,        % color of file links
    urlcolor=blue,        % color of external links
    linktoc=page          % only page is linked
 }
\begin{document}

%\preprint{APS/123-QED}
\large
\title{\Large Inducing half metallicity with alloying in Heusler Compound CoFeMnSb}% Force line 

\author{Upendra Kumar, P. V. Sreenivasa Reddy, Satadeep Bhattacharjee, Seung-Cheol Lee}
 \altaffiliation[]{Indo-Korea Science and Technology Center (IKST), New Airport Road, Yelahanka, Bangalore, India- 560065.}%Lines break automatically or can be forced with \\
\date{\today}% It is always \today, today,
\email{seungcheol.lee@ikst.res.in}

\begin{abstract}
\noindent  First principles studies were performed in order to find out the possibility of inducing half-metallicity in Heusler Compound CoFeMnSb, by means of alloying it with 3d-transition metal elements. Proper alloying element is selected through the calculations of formation energies. These calculations were tested with different concentrations of alloying elements at different atomic sites. Among the selected transition metal elements Sc and Ti are proposed to be excellent alloying elements particularly at Mn site. By using these alloying elements complete half metallic behaviour is obtained in $\mathrm{CoFeMn_{0.25}Sc_{0.75}Sb}$, $\mathrm{CoFeMn_{0.75}Ti_{0.25}Sb}$, $\mathrm{CoFeMn_{0.625}Ti_{0.375}Sb}$,  $\mathrm{CoFeMn_{0.50}Ti_{0.50}Sb}$, $\mathrm{CoFeMn_{0.25}Ti_{0.75}Sb}$ and CoFeTiSb alloys. Shifting of Co-Fe d-states towards lower energy region leads to zero density of states at Fermi level for the spin minority channel. Alloying effects on the electronic structure and magnetization are discussed in details. Thermodynamical  stability of these new alloys are major part of this study. The Curie temperatures of $\mathrm{CoFeMn_{0.25}Sc_{0.75}Sb}$ and $\mathrm{CoFeMn_{0.75}Ti_{0.25}Sb}$  were found to be 324.5 K and 682 K; respectively, showing good candidature for spintronics applications. For understanding the bonding nature of constituent atom of CoFeMnSb, crystal orbital Hamiltonian populations have been analysed. 


%\begin{description}
%\pacs{31.15.E}
%\end{description}
\end{abstract}
%\keywords{Atom beam experiment; qubit manipulation; Rb atom beam}
%\pacs{42.50.Ex, 32.80.Wr, 32.80.-t; 32.10.Fn}
%% PACS, the Physics and Astronomy
%                             % Classification Scheme.
\maketitle
%\tableofcontents
\onecolumngrid

\section{ Introduction}
The Heusler class of materials are well known for their plenty of applications ranging from spintronics to thermoelectric and superconductors to topological insulators due to their wide range of material properties that can be engineered. 
These Heusler compounds seem to be the material of choice for many applications because of their tunable electronic structure properties. This helps in the design of materials with desirable properties from half-metallic ferromagnet  \cite{Kubler,Felser} over completely ferrimagnet \cite{Wurmehl} to non-magnetic semiconductors  \cite{Jung,Pierre}. In the present days, the search of new materials for spintronics applications was growing within the Heusler class of compounds which was started before 30 years  \cite{Kubler}. These materials have passionate applications in spintronics \cite{Groot,Felser1,Felser2}, optoelectronics \cite{Kieven}, as shape memory \cite{Christian} alloys and as superconductors \cite{Wernick,Winterlik}. Over the past few decades, new areas of applications emerged comprising the environmental technologies such as thermoelectric \cite{Sakurada,Uher} and solar cell applications \cite{Kieven}. Recently, a new trend in the material properties which belong to new quantum state of matter is observed \cite{Chadov,Lin}. These materials are named as topological insulators and are having technological importance due to the conduction of electrons being extremely high because of its surface states and might attract application in high performance electronic systems. This could be one of the future research perspective of Heusler compounds enabling them to be more productive.

Ternary intermetallic Heusler compounds can be represented with general formula XYZ (known as Half-Heusler) or X$_2$YZ, where X and Y elements generally belong to transition metals (having $d$ states) and Z belongs to main group element. In 1903 Fritz Heusler  \cite{Heusler,Heusler1} discovered these class of materials. Important feature of these compounds is, the properties of the total system is completely different from the properties of the individual elements contained in this type of compounds. For example, if we consider the compound Cu$_2$MnAl, which is reported as magnetic none of its constitutional elements possess magnetic order independently. Similarly, in another compound TiNiSn, which is reported to have semiconducting nature all its constitutional elements posses metallic character  \cite{Aliev}. The count on the number of these type of compounds is endless and can be prepared by mixing all the elements in the periodic table and leading to several applications and need to be explored.

By doping in Heusler compounds, the Fermi level can be adjusted and half metalicity can be increase as in the case of Co$_2$MnGe$_{1-x}$Ga$_{1-x}$ \cite{Wu}. The co-doping effect of Co+Si on nearly half-metallic Heusler compound NiFeMnSn has been seen and shown that by doping it becomes perfectly half-metallic \cite{Feng}. The more $d$-electrons doping effects has been seen in Ti$_2$CoAl by Chen et al. \cite{Chen}, it was shown how V/Nb doping is more energetically favourable on different site of Ti. Effect of Fe substitution by Co on off-stoichiometric Ni-Fe-Co-Mn-Sn Heusler alloy ribbons has been seen and it was predicted that Fe substitution by Co creates change in the magnetic behavior \cite{Mishra}.

Considerable attention had paid in the quaternary Heusler compounds like CoFeMnAl, CoFeMnSi, CoFeMnGa, CoFeMnGe, CoFeScSb, CoFeTiSb, NiFeMnGa and NiCoMnGa. There are two types of half-metallic materials: the complete half-metals with 100\% spin polarization at the Fermi-level and the near half-metals where there is some small density of states at the Fermi level so that the spin polarization is less than 100\%  \cite{Kundu}. The compounds like CoFeMnSi, CoFeMnAs and CoFeTiSb \cite{Berri} are belongs to the 100\% spin polarization category. But CoFeMnSb have only 85\% spin polarization \cite{Elahmar} and CoFeScSb is reported to be nearly half metallic \cite{Gao}. It is therefore worth to investigate the possibility of 100\% spin polarization in CoFeMnSb by means of fractional alloying. The study of mechanical, structural, electronic, magnetic and thermoelectric properties of CoFeMnSb ternary Heusler alloy has already been done by spin polarized density functional and semiclassical Boltzmann transport theory  \cite{Lee}. It was found, CoFeMnSb is more thermodynamically stable in ferromagnetic phase  \cite{Lee}. %It was found that CoFeMnSb is nearly half-metallic with spin polarization 85$\%$  \cite{Elahmar}. 

Calculation of formation energy plays a key role in identifying the thermodynamic stability of a predicted or alloyed compound. The negative value of formation energy indicates that at zero temperature the compound is more stable than its constituent elements. By using the formation energy criteria Ma  et al. discovered 45 half-metals and 34 nearly half metals by identifying the negative formation energies \cite{ma2017computational}. Understanding the nature of bonding is another important task, which can be performed by analysing the {\it Crystal Orbital Hamilton Population} (COHP) scheme  \cite{Dronskowski} in which partitioning of the energies allows to extract chemical bindings between the atoms from the electronic structure calculations. This can be done with LOBSTER package \cite{Stefan Maintz} which is, in general, used to extract chemical bonding from plane-wave based density functional theory.

In this article, we have theoretically investigated the alloying element effect of transition metal (TM) (TM= Sc, Ti, V, Cr, Ni, Cu, Zn) on CoFeMnSb ternary Heusler alloy by using  density functional theory (DFT) and see how half metalicity of CoFeMnSb changes with alloying element. Thermodynamic stability of these new alloys are major part of our study. It was found Sc and Ti alloying is more effective and change nearly nearly half-metallic nature of CoFeMnSb to completely half-metallic. The paper is organized as follows: section \ref{CD} presents computational details of the first-principles calculations. The results and discussions of bulk ground state, electronic structure, formation energy calculations and electronic structure details of alloyed CoFeMnSb are presented in section \ref{RD}. Finally, conclusions are presented in the section \ref{Conclusions}.

\section{ Computational details}
\label{CD}
All the ground state and electronic structure calculations are performed by using Vienna ab-initio simulation package (VASP) \cite{vasp, vasp1}. Firstly, structural optimizations for both bulk and alloyed structures were performed to obtain the minimal energy structure using the Perdew-Burke-Ernzerhof (PBE) functional \cite{PBE}, which is specifically customized for solids, that has been shown to yield structural data in agreement with experiment. The PBE pseudopotential was used with an energy cutoff of 600 eV. The Brillouin zone has been sampled using a 8$\times$8$\times$8 $\Gamma$-centred grid. The criteria for energy convergence between the two successive self consistent loop is taken $10^{-6}$ eV/cell. We have used tetrahedron method with Bl{\"o}chl correction  \cite{Bloch}, along with 8$\times$8$\times$8 K-point mesh for the density of states calculations. There is fully relaxation of atomic structures and volume by using convergence gradient algorithm. Forces are converged to 0.01 eV/$\AA$. For study of alloying in CoFeMnSb, a super cell of 2$\times$2$\times$2 has been designed from the primitive cell which contains 8 atoms of each kind. In next section, we have discussed the effects of TM alloying in CoFeMnSb on Co, Fe, Mn sites and how these alloying elements affect half- metalicity and magnetism.

\section{ Results and Discussion}
\label{RD}
Present compound CoFeMnSb crystallizes in face centred cubic type with space group $F\overline{4}3m$ (216) and the unit cell occupied with the Wyckoff positions: Co (0 0 0), Fe (0.5 0.5 0.5), Mn (0.25 0.25 0.25) and Sb (0.75 0.75 0.75). We have created a 2$\times$2$\times$2 supercell which is consisting of total 32 atoms. To start with, we performed complete structural optimization for the supercell created in spin polarization case and the calculated equilibrium lattice constant is found to be 8.44$\AA$. The calculated total magnetic moment of the system is observed as 5.005$\mu$$_B$ per primitive cell. The atom dependent magnetic moments are observed as 0.894$\mu$$_B$ for Co, 1.071$\mu$$_B$ for Fe, 3.077$\mu$$_B$ for Mn and -0.037$\mu$$_B$ for Sb. From this it is observed that Mn, Fe and Co elements are contributing positively to the total magnetic moment while Sb is providing negative contribution. It is also observed that the primary contribution to the total magnetic moment is arrived from the Mn and the secondary contribution is arriving from Fe and Co sites.

\begin{figure}[H]
\centering
\subfigure[]{\includegraphics[width=60mm,height=50mm]{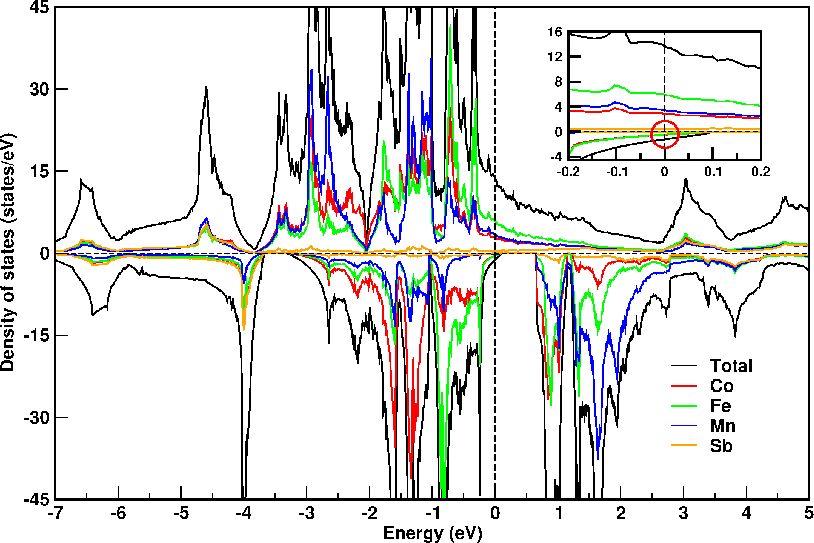}}
\subfigure[]{\includegraphics[width=60mm,height=50mm]{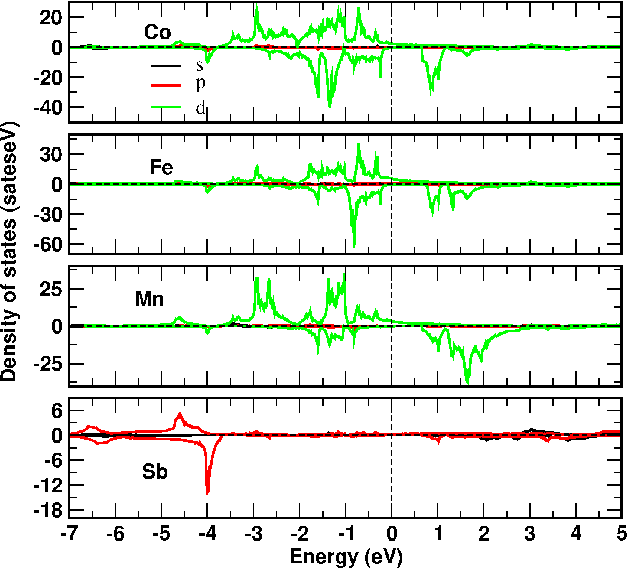}}
\caption{(a) Total and atom projected electronic density of states for bulk CoFeMnSb in 2$\times$2$\times$2 supercell case. (b) Orbital projected density of states for atom.}
\label{bulk-dos}
\end{figure}

To know more about the electronic structure details, we have calculated total and atom projected electronic density of states and are plotted in Fig.\ref{bulk-dos}. Multiple peaks are observed in the total density of states both in majority and minority spin cases. The peak around -4.6 eV in majority spin and -4 eV in minority spin are due to the hybridization of all elements. Particularly at -4 eV it is due to the Sb-$p$ states. The peaks around -3.0 eV to -2.5 eV and -1.4 eV to -1 eV are mainly due to Mn atoms in the majority spin case. The states in the energy region form -1.0 eV to 0 eV in both majority and minority are highly dominated by Fe atom. For Mn atom the split in the $d$ states is more in compare to Fe and Co, which causes for the high magnetic moment in Mn in compare to Fe and Co in the present system. From the overall plot the system is found to nearly half metallic with large density of states in majority spin case and very small value of density of states in minority spin case at Fermi level (E$_F$). So, it is quite informative to discuss about the system near the Fermi level. From the inset of Fig.\ref{bulk-dos}, at E$_F$, the decreasing trend in the density of states observed from the contribution of Fe, Mn, Co and Sb atoms respectively in majority spin case but in the minority spin case we have observed a small value of density of states which are arising mainly from Co and Fe atoms (indicated with red color circle). As the present compound is almost having a half metallic nature it is worth to discuss about the spin polarization, $P$, in the present compound. The well known formula to calculate $P$ is $P=((N(E_F)\uparrow-N(E_F)\downarrow)/(N(E_F)\uparrow+N(E_F)\downarrow))\times 100$\%, where $N(E_F)\uparrow$ is total density of states at E$_F$ in majority spin case and $N(E_F)\downarrow$ is total density of states at E$_F$ in minority spin case. The calculated value of $N(E_F)\uparrow$ is 13.69 states/eV and $N(E_F)\downarrow$ is 1.229 states/eV. From this, the estimated $P$ is 83.52\% which is in good agreement with other studies (85\%) \cite{Elahmar}. As the present compound is found to nearly half metallic, it could be more useful if we can tune the Fermi level to get complete half metallic nature with 100\% spin polarization. The best way to tune the Fermi level is fractional alloying with TM elements in the periodic table. 

\subsection{ Proper selection of alloying element from the calculation of formation energies}
In order to know the possibility to synthesize a compound, calculation of formation energy ($\Delta E_{f}$) is the best way. The formation energy can be written as, $\Delta E_{f}=E^{'}-E-\sum_{i}n_{i}\mu_{i}$, where $E^{'}$ is the total energy of the supercell containing the impurity atom, $E$ represents the total energy of the perfect crystal using an equivalent supercell, integer $n_i$ is the number of alloying element atoms, which has been subtracted (negative value) or added (positive value) to form the disorders and $\mu_{i}$ represents the chemical potential of these species. The calculated value of $\Delta E_{f}$ for bulk CoFeMnSb is found to be -0.68 eV/f.u. So, the alloyed compound should have the lower values of $\Delta E_{f}$ in compare to bulk. 

\begin{figure}
\vspace{-1.5cm}
\subfigure[]{\includegraphics[width=70mm,height=65mm]{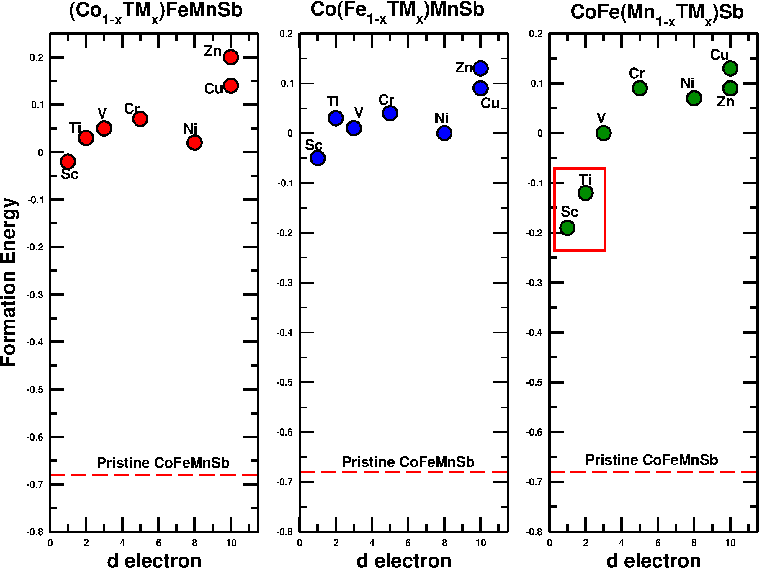}}
\subfigure[]{\includegraphics[width=70mm,height=65mm]{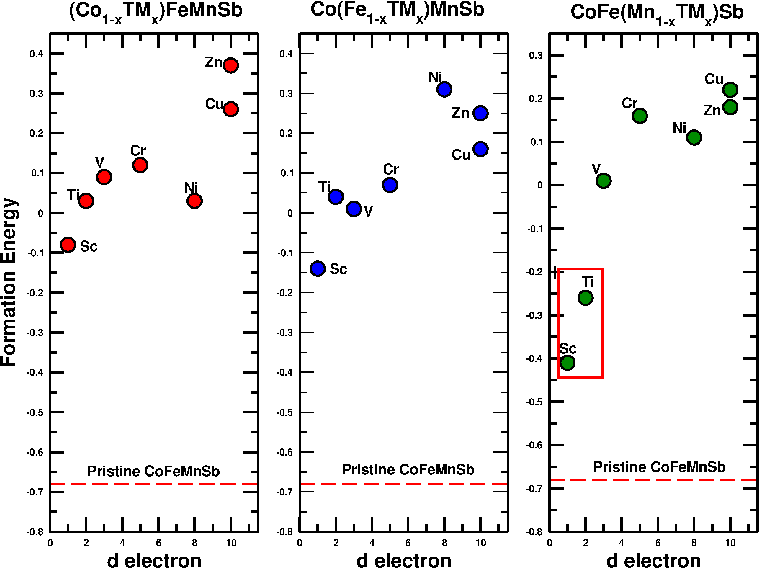}}
\subfigure[]{\includegraphics[width=70mm,height=65mm]{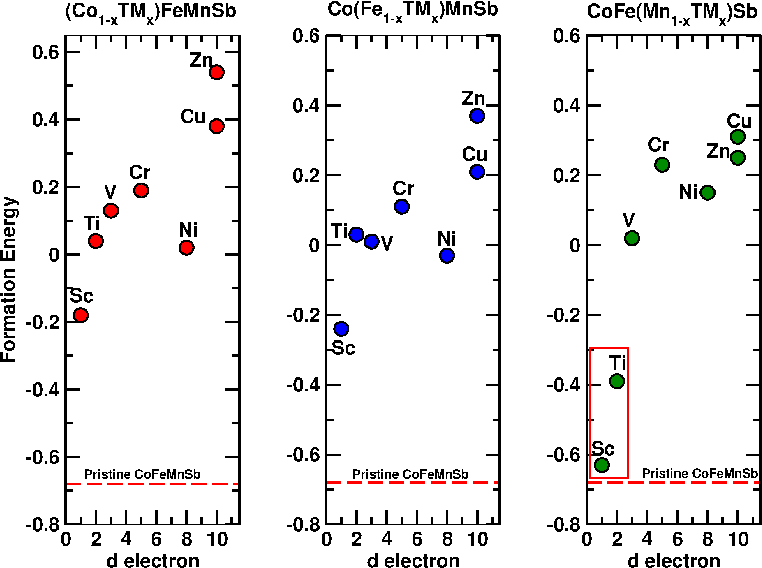}}
\subfigure[]{\includegraphics[width=70mm,height=65mm]{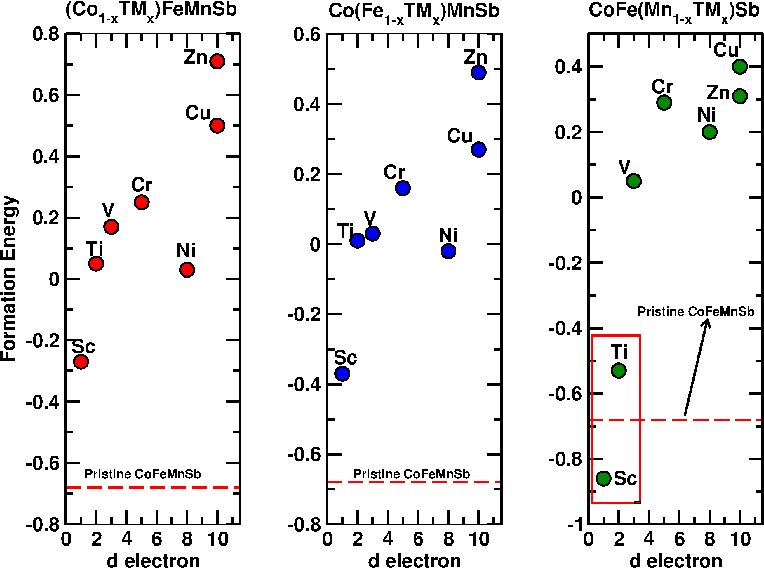}}
\subfigure[]{\includegraphics[width=70mm,height=65mm]{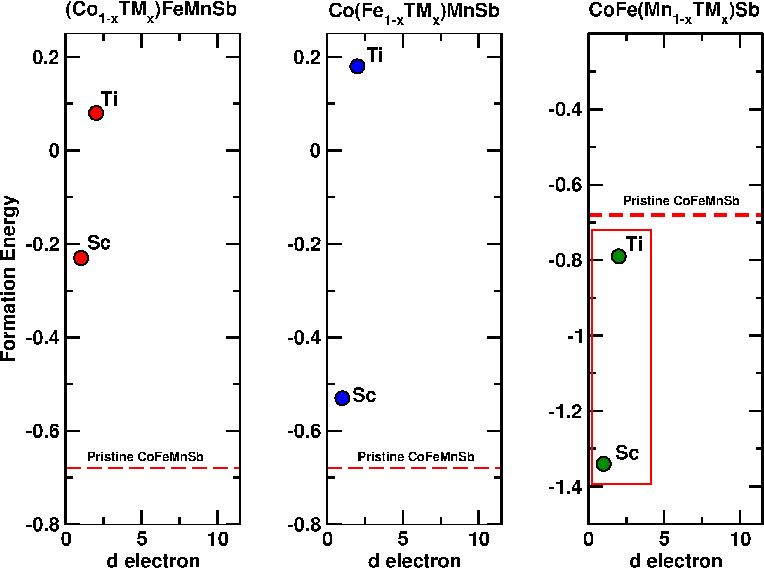}}
\subfigure[]{\includegraphics[width=70mm,height=65mm]{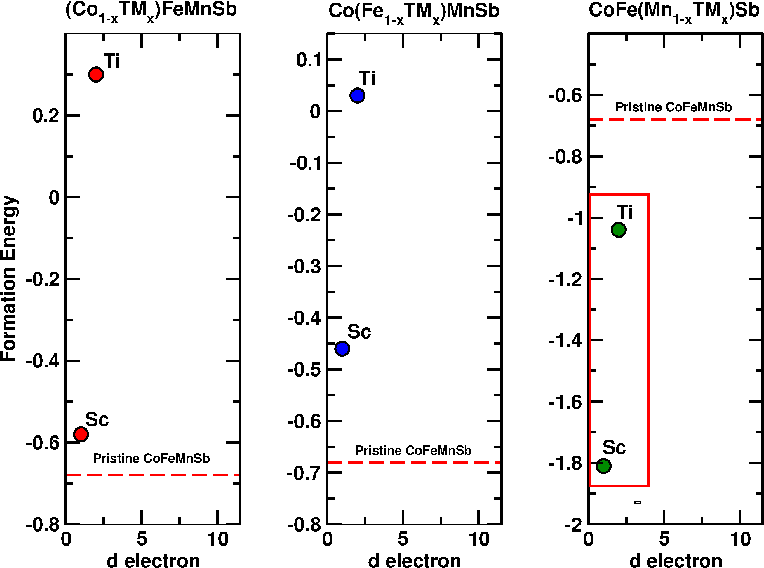}}
\caption{Calculated formation energy (in eV) with alloying concentration (a) $x$=0.125, (b) $x$= 0.250, (c) $x$= 0.375, (d) $x$= 0.500, (e) $x$= 0.750 and (f) $x$= 1.00 at Co, Fe and Mn sites.}
\label{alloyant_concentration}
\end{figure}

Selection of a proper alloying element is always a difficult task. It includes lot of trails with different alloying elements along with different concentrations at different atomic sites. In the present case, we have selected several TM elements as alloying elements to alloy at Co, Fe and Mn sites. The selected TM elements are Sc, Ti, V, Cr, Ni, Cu and Zn. We have tested these alloying elements at Co, Fe and Mn sites with different alloying element concentrations, $x$= 0.125, 0.25, 0.375, 0.50, 0.75 and 1.00 to check the site preference for a particular alloying element. The calculated formation energy values are plotted in Fig.\ref{alloyant_concentration}. From this it is observed that Sc alloying element has lower formation energies in compare to other elements with all alloying concentrations and all Co, Fe and Mn site preferences which confirms that Sc could be the best alloying element among the chosen elements. Along with Sc, Ti could be the second possible alloying element at Mn site with all alloying concentrations. These results are highlighted with a rectangular red coloured box for all $x$ concentrations at Mn site alloying. It is also observed that, as the alloying elements concentration $x$ increases from 0.125 to 0.375, the formation energy of CoFe(Mn$_{1-x}$TM$_x$)Sb is approaching the bulk formation energy value which is indicated with red coloured dotted line. At $x$= 0.500 alloying concentration of Sc, the formation energy of $\mathrm{CoFeMn_{0.50}TM_{0.50}Sb}$ is lowering than the bulk indicating best choice of alloying concentration of Sc in the formation of alloy. As $x$ concentration value increases further 0.75 and 1.00 both Sc and Ti alloyed CoFe(Mn$_{1-x}$TM$_x$)Sb having lower $\Delta E_{f}$ values in compare to bulk indicating Ti will be the second possible element to synthesize CoFe(Mn$_{1-x}$TM$_x$)Sb alloy after Sc.

\subsection{ Alloyant effect on electronic structure and magnetic nature} 

To investigate the effect of alloying on the electronic structure of CoFeMnSb, we have calculated total electronic density of states for $x$= 0.125, 0.250, 0.375 and 0.500 concentrations alloyed with TM and are plotted in Fig.\ref{alloyant_dos}. In all the cases, majority spin states of CoFeMnSb in the presence various alloying elements has been stride over the Fermi level, describing typical metallic nature. On the other hand, the minority spin states, gradually decreases and becomes zero, describing half metallic nature of CoFeMnSb in presence various alloying elements along with alloying elements percentage. From Fig.\ref{alloyant_dos}, half metallic nature is observed in (Co$_{1-x}$Sc$_x$)FeMnSb with $x$=0.125, 0.25, 0.375 and 0.50, in Co(Fe$_{1-x}$Sc$_x$)MnSb with $x$= 0.125, 0.25 and 0.375, in $\mathrm{Co_{0.875}Ti_{0.125}FeMnSb}$, CoFe(Mn$_{1-x}$Sc$_x$)Sb with $x$= 0.125, 0.250, 0.375 and 0.500 is almost half metallic, in $\mathrm{Co_{0.875}Ti_{0.125}FeMnSb}$, in CoFe(Mn$_{1-x}$Ti$_x$)Sb with $x$= 0.250, 0.375 and 0.500, in CoFe(Mn$_{1-x}$V$_x$)Sb with $x$= 0.125, 0.250, 0.375 and 0.500, in Co(Fe$_{1-x}$Cr$_x$)MnSb with $x$= 0.250, 0.375 and 0.500, in $\mathrm{CoFeMn_{0.50}Cr_{0.50}Sb}$, in (Co$_{1-x}$Ni$_x$)FeMnSb with $x$= 0.125, 0.250, 0.375 and 0.500, in $\mathrm{CoFe_{0.875}Ni_{0.125}MnSb}$, in CoFe(Mn$_{1-x}$Ni$_x$)Sb with $x$= 0.125 and 0.250, in $\mathrm{Co_{0.875}Cu_{0.125}FeMnSb}$, in Co(Fe$_{1-x}$Cu$_x$)MnSb with $x$= 0.375 and 0.500, in $\mathrm{CoFeMn_{0.875}Cu_{0.125}Sb}$, in $\mathrm{Co_{0.875}Zn_{0.125}FeMnSb}$, in Co(Fe$_{1-x}$Zn$_x$)MnSb with $x$= 0.125, 0.250, 0.375 and 0.500. 

\begin{figure}
\centering
\includegraphics[width=60mm,height=55mm]{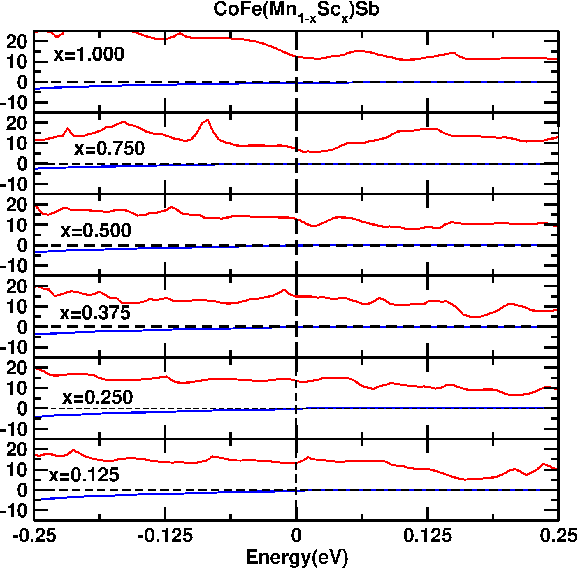}
\includegraphics[width=60mm,height=55mm]{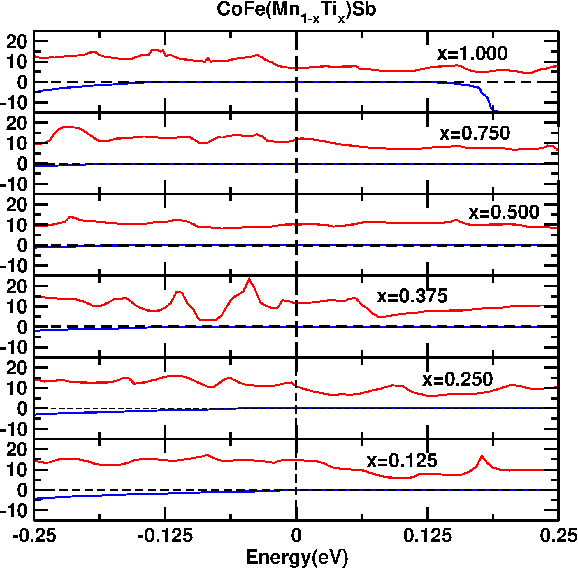}
\caption{Doping effect on electronic structure of CoFeMnSb at Mn site with Sc and Ti.}
\label{alloyant_dos1}
\end{figure}

\begin{figure}
\centering
\includegraphics[width=60mm,height=55mm]{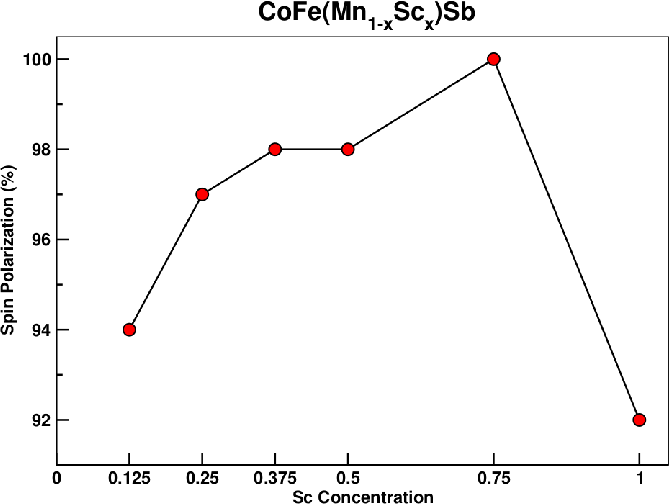}
\includegraphics[width=60mm,height=55mm]{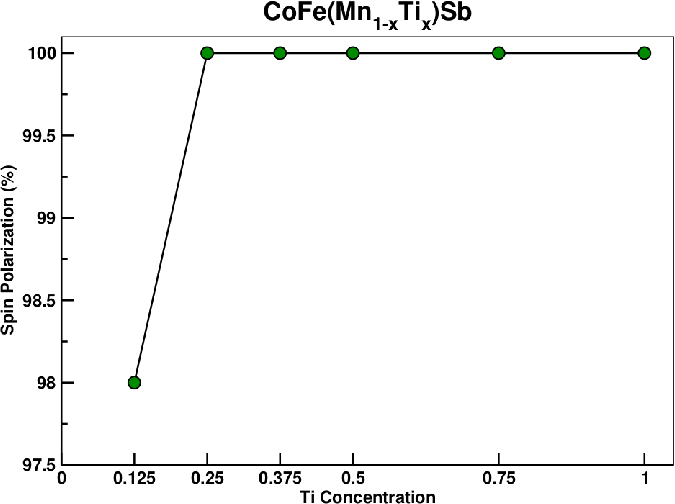}
\caption{Spin polarization for Sc and Ti alloying at Mn site }
\label{alloyant_polarization}
\end{figure}

Even though we have observed half metallic nature with different alloying concentrations, their formation energies are not lower than the bulk one. From the previous subsection we have found only Sc and Ti are the two possible alloying elements particularly at Mn site. So it is worth to discuss about only Sc and Ti alloying studies at Mn site. Calculated total density of states are plotted separately for Sc and Ti alloying elementss only at Mn site with different alloying concentrations in Fig.\ref{alloyant_dos1}, where we can observed complete half metalicity in $\mathrm{CoFeMn_{0.25}Sc_{0.75}Sb}$, $\mathrm{CoFeMn_{0.75}Ti_{0.25}Sb}$, $\mathrm{CoFeMn_{0.625}Ti_{0.375}Sb}$,  $\mathrm{CoFeMn_{0.50}Ti_{0.50}Sb}$,\\ $\mathrm{CoFeMn_{0.25}Ti_{0.75}Sb}$ and CoFeTiSb. The calculated band structures are given in Fig.\ref{band} for bulk, $\mathrm{CoFeMn_{0.25}Sc_{0.75}Sb}$, $\mathrm{CoFeMn_{0.50}Ti_{0.50}Sb}$ and CoFeTiSb to know the type of band gap in minority spin case. In all these cases we have observed a direct gap, where top of the valence band and bottom of conduction band are on the $\Gamma$ high symmetry point. This is completely different form the half metallic C1$_b$ Heusler compounds which are normally having an indirect gap. Calculated spin polarization values for Sc and Ti alloying element at Mn site are given in Fig.\ref{alloyant_polarization} along with different alloying concentrations. In Sc alloying at Mn site, the spin polarization value is found to increase with alloying concentration and observed a 100\% spin polarization at $x$=0.75. At the higher alloying concentrations of Sc greater than $x$=0.75, spin polarization is found to decrease. From the literature, CoFeScSb is found to be nearly half-metallic \cite{Gao} which is again confirmed from our calculations by complete replace of Mn with Sc. Here we can conclude one important point is that if we replace 0.25\% of Sc with Mn in CoFeScSb we will get 100\% spin polarization. In the case of Ti alloying at same Mn site, 100\% spin polarization is observed at $x$= 0.25 and above concentrations. From the literature CoFeTiSb is found to have complete half-metallic \cite{Berri} which is again confirmed from our calculations.

To know the effect of magnetic nature with alloying, we have calculated total and atom dependent magnetic moments for both Sc and Ti alloying elementss with different concentrations and are tabulated in Table.\ref{magnetic}. It is observed that total and atom dependent magnetic moments are found to be decreased with increasing alloying concentration in the case of Sc at Mn site. The same scenario is observed with total, Sb and alloying element magnetic moments with Ti at Mn site but Co, Fe and Mn moments are found to increased a little with increasing Ti alloying concentration at Mn site.

\begin{figure}
\centering
\subfigure[]{\includegraphics[width=50mm,height=58mm]{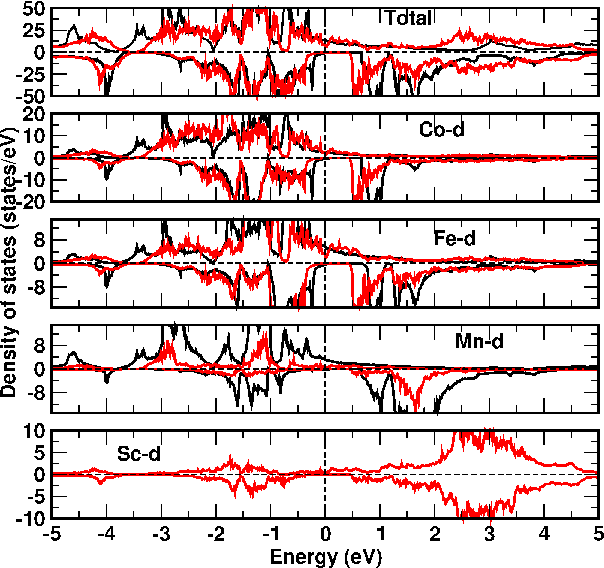}}
\subfigure[]{\includegraphics[width=50mm,height=58mm]{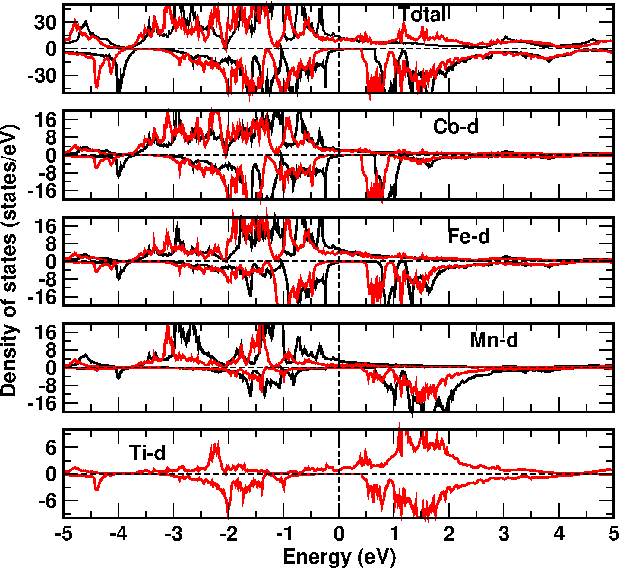}}
\subfigure[]{\includegraphics[width=50mm,height=58mm]{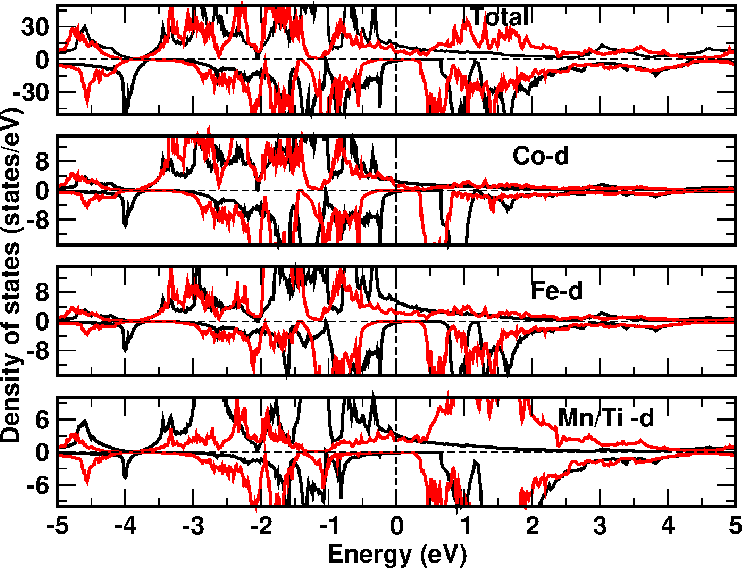}}
\caption{Total and $d$ orbital projected density for (a) without alloying and with alloying of 0.75\% Sc at Mn site, (b) without alloying and with alloying of 0.50\% Ti at Mn site and (c) without alloying and with alloying of 100\% Ti at Mn site. In all plots black color data indicates without alloying and red color data indicates with alloying at Mn site. }
\label{dope_dos}
\end{figure}

Change in the electronic structure with alloying can be understand through the electronic density of states which are plotted in Fig.\ref{dope_dos} for Sc with 75\% alloying, Ti with 50\% and 100\% alloying along with bulk states. In all cases, the total density of states in minority spin case are found to be shifted towards to the lower energies which causes a gap at the Fermi level in minority spin. The states near E$_F$ in minority spin case mainly belongs to $d$ states of Co and Fe in bulk. These states are found to be shifted towards the lower energies and cause a 100\% spin polarization with alloying. It is also observed reduction in the height of peaks in Mn $d$ states after alloying with Sc or Ti. After alloying, the splitting of Sc and Ti $d$ states is very less as compare to Mn in bulk which causes a decreasing trend in the total magnetic moment with alloying compare to bulk value.

To obtain the Curie temperature, we have used the SPR-KKR-method  \cite{SPR-KKR}, by which we have calculated the inter-atomic exchange constants. The Curie temperatures were calculated within the mean field approximation. It was found CoFeMnSb has Curie temperature 923.1 K, which decreases by including alloying concentration at Mn site (described in Table \ref{table}). CoFeMnSb and CoFeScSb are not a perfect half metal, have spin polarization 83.52\% and 92\% respectively. But $\mathrm{CoFeMn_{0.25}Sc_{0.75}Sb}$ has complete half metallic nature and Curie temperature 324.5 K i.e. more than room temperature. So $\mathrm{CoFeMn_{0.25}Sc_{0.75}Sb}$ is a perfect Heusler compound for spintronics application. The Curie temperature of $\mathrm{CoFeMn_{0.75}Ti_{0.25}Sb}$ is found 682 K, also has a possibility in spintronics application. This shows that we can engineer half-metallic compounds by alloying CoFeMnSb with Ti.

\begin{figure}
\centering
\includegraphics[width=120mm,height=100mm]{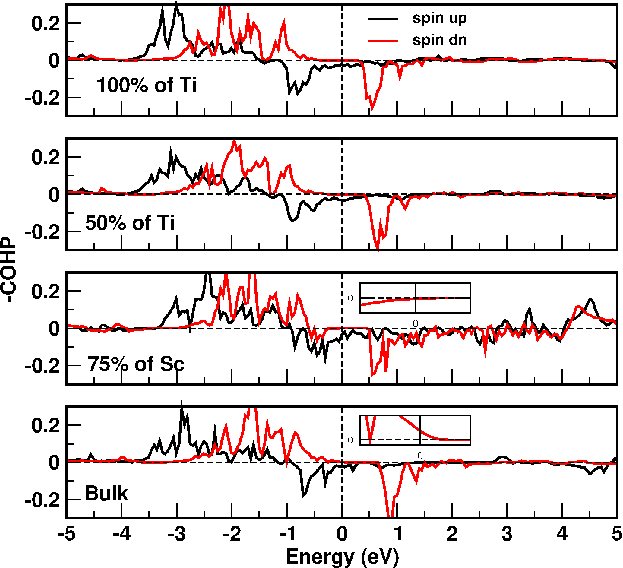}
\caption{Crystal orbital Hamiltonian population analysis for Co-Fe.  }
\label{COHP}
\end{figure}

Above discussions intimating the importance of Co and Fe $d$ states, which are highly effected with alloying and reason for half metallic nature after alloying. So, it is very important to understand the bonding nature in between Co and Fe, and how it is effected with alloying. For this we have computed Co-Fe crystal orbital Hamiltonian populations(COHPs) \cite{Dronskowski}. COHP is the density of states weighted by the corresponding Hamiltonian matrix element. COHPs indicate the strength as well as the nature of interaction with the bonding or antibonding interactions. These interactions being indicated by positive or negative values of COHP. Co-Fe COHPs are plotted in Fig.\ref{COHP} for bulk, $\mathrm{CoFeMn_{0.25}Sc_{0.75}Sb}$, $\mathrm{CoFeMn_{0.50}Ti_{0.50}Sb}$ and CoFeTiSb. It is observed that bonding states in both spin cases at energies below around -1 eV. From the energy range -1 eV to 0 eV we have antibonding states in majority spin case for both bulk and new alloys. In the same energy range, the scenario is different in minority spin case, where we have observed bonding states in bulk and Ti alloying but in Sc alloying we have antibonding nature. At the E$_F$, we have small electronic density of states in minority spin case without alloying are belongs to Co and Fe. The inset of the figure indicates that these states are having bonding nature. This bonding nature still survived with Ti alloying but it changes to antibonding nature with Sc alloying. 

\section{ Conclusions}        
\label{Conclusions}
We have investigated the TM element alloying effect on the electronic structure and magnetic properties of nearly half metallic CoFeMnSb by using first principles calculations within density functional theory. Calculations on the relative formation energy indicates Sc and Ti mono alloying is more favourable. The same calculations confirms the best site preference to alloy with Sc and Ti is at Mn-site. Therefore, the mono alloyed CoFe(Mn$_{1-x}$Sc$_{x}$)Sb and CoFe(Mn$_{1-x}$Ti$_{x}$)Sb alloys could be easily obtained. The calculated electronic structures reveals 100\% spin polarization when alloying concentration, $x$, reaching to $x$=0.75 with Sc and $x$= 0.25, 0.375, 0.50, 0.75, 1.0 with Ti at Mn site. $\mathrm{CoFeMn_{0.25}Sc_{0.75}Sb}$ and $\mathrm{CoFeMn_{0.75}Ti_{0.25}Sb}$ are perfectly half metallic with 100\% spin polarization, have Curie temperature 324.5 K and 682 K, respectively. Therefore, both become good candidature in spintronics application. Density of states and COHP calculations confirms that both Co and Fe states are shifted towards to lower energy region and causes to open a gap at E$_F$ in the minority states. This leads to direct band gap in the minority spin states in these materials.

\FloatBarrier
\begin{table}[H]
\caption{\large Magnetic moments and Curie temperature for different alloying concentrations of Sc and Ti at Mn site.}
\label{magnetic}
%\begin{ruledtabular}
\begin{adjustbox}{width=\columnwidth,center}
\scalebox{0.5}{
\begin{tabular}{cccccccccc}
alloying elements &$x$		&$\mu$$_{B(Co)}$	&$\mu$$_{B(\mathrm{alloying\; element})}$	&$\mu$$_{B(Fe)}$ &$\mu$$_{B(Mn)}$	&$\mu$$_{B(Sb)}$	&$\mu$$_{B(Total)}$ & T$_\mathrm{c}$(K)\\
\hline				
Sc	&0.000		&0.894	&0.0	&1.071	&3.077	&-0.037	&5.005&923.1\\
	&0.125		&0.869	&-0.194	&1.018	&3.054	&-0.031	&4.504&-\\
	&0.250		&0.839	&-0.185	&0.960	&3.049	&-0.026	&4.013&-\\
	&0.375		&0.815	&-0.168	&0.889	&3.048	&-0.020	&3.526&-\\
	&0.500		&0.796	&-0.162	&0.823	&3.032	&-0.015	&3.039&-\\
	&0.750		&0.719	&-0.153	&0.768	&2.792	&-0.006	&2.064&324.5\\
	&1.00		&0.514	&-0.143	&0.738	&0.00	&-0.010	&1.099&222.0\\
\hline
Ti	&0.00		&0.894	&0.0	&1.071	&3.077	&-0.037	&5.005&923.1\\
	&0.125		&0.896	&-0.362	&1.094	&3.084	&-0.031	&4.612&-\\
	&0.250		&0.898	&-0.330	&1.110	&3.114	&-0.025	&4.236&682.0\\
	&0.375		&0.905	&-0.289	&1.118	&3.150	&-0.019	&3.864&-\\
	&0.500		&0.908	&-0.246	&1.137	&3.166	&-0.012	&3.493&-\\
	&0.750		&0.924	&-0.176	&1.213	&3.211	&-0.002	&2.805&-\\
	&1.00		&1.017	&-0.203	&1.204	&0.00	&0.009	&2.027&-\\
\hline
\end{tabular}}
%\end{ruledtabular}
\end{adjustbox}
\label{table}
\end{table}


\clearpage
\section*{Supplementary Material}
%%%%%%SUPPLEMENT MATERIAL %%%%%%%%%%%%%%%
\FloatBarrier
\newcommand{\beginsupplement}{%
        \setcounter{table}{0}
        \renewcommand{\thetable}{S\arabic{table}}%
        \setcounter{figure}{0}
        \renewcommand{\thefigure}{S\arabic{figure}}%
     }

\beginsupplement

\begin{figure}
\centering
  \begin{tabular}{@{}cccc@{}}
\includegraphics[width=50mm,height=50mm]{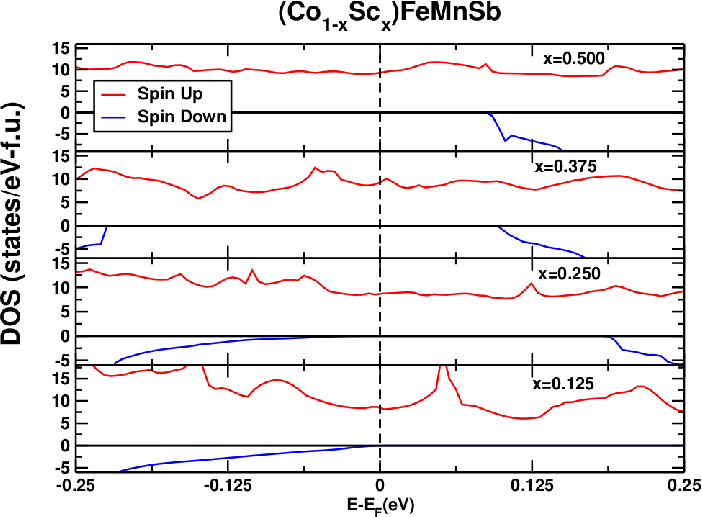}
\includegraphics[width=50mm,height=50mm]{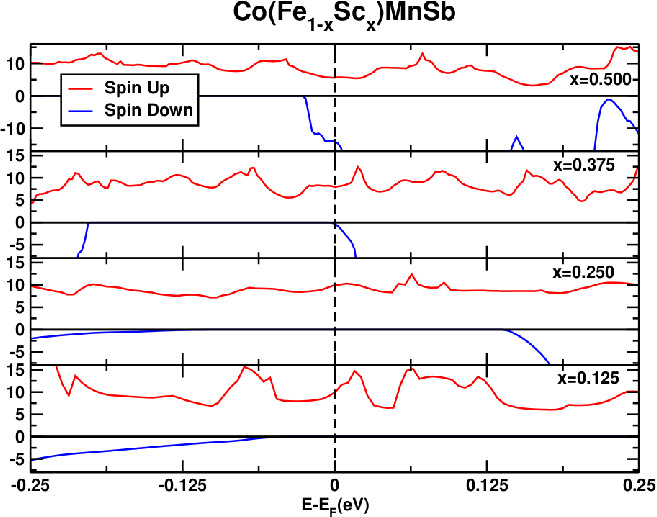}
\includegraphics[width=50mm,height=50mm]{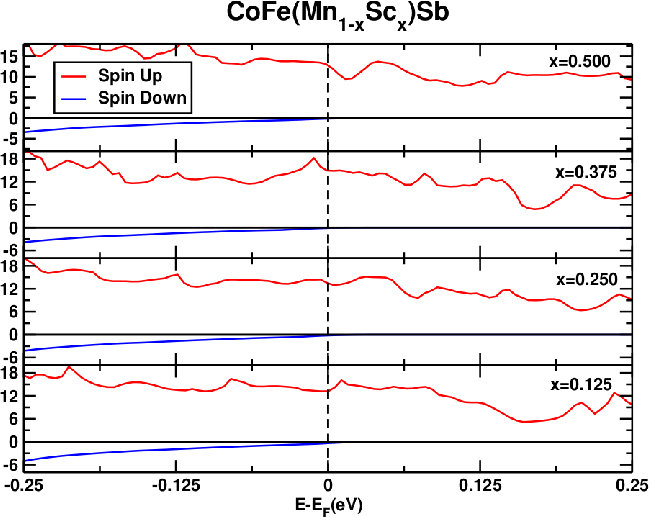}\\
\includegraphics[width=50mm,height=50mm]{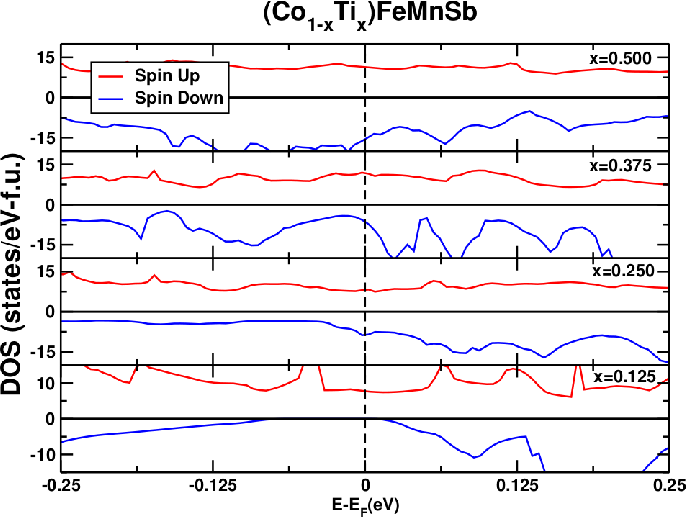}
\includegraphics[width=50mm,height=50mm]{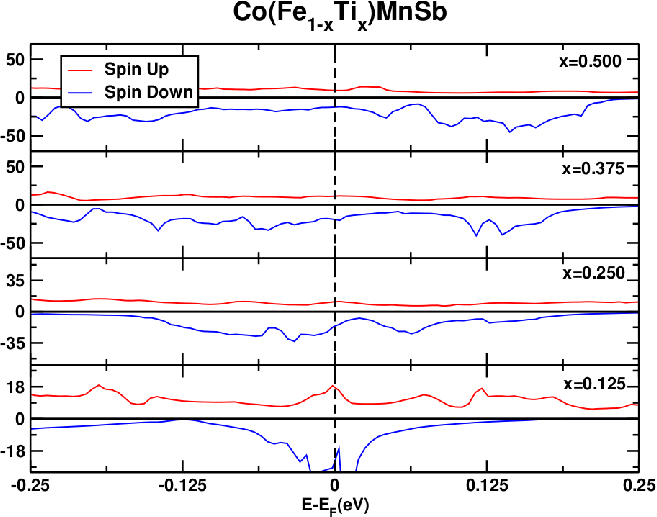}
\includegraphics[width=50mm,height=50mm]{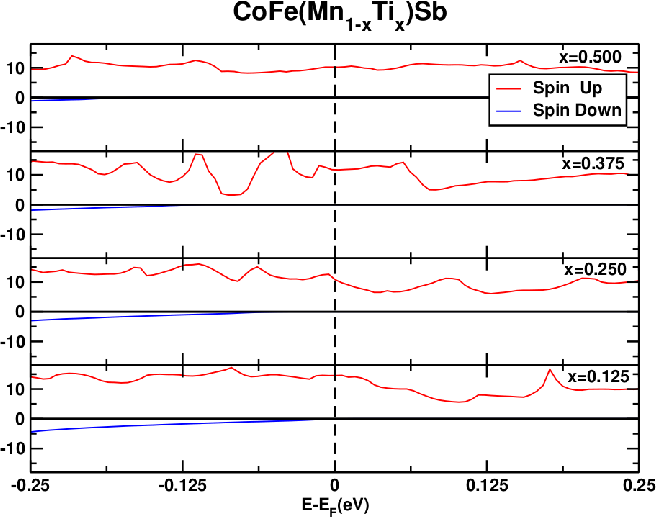}\\
\includegraphics[width=50mm,height=50mm]{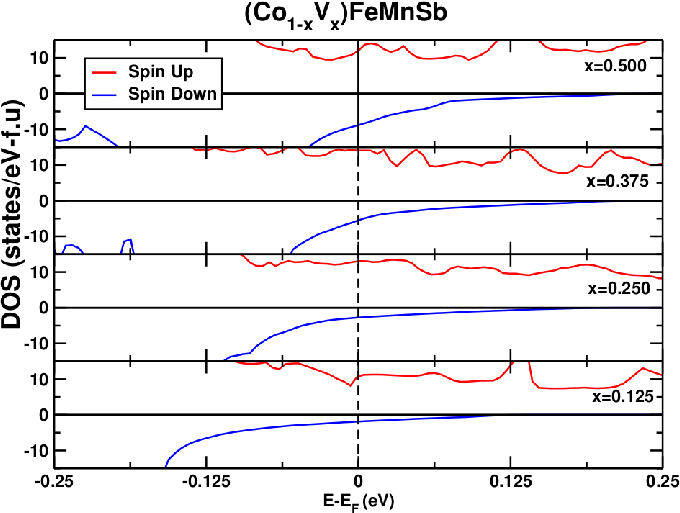}
\includegraphics[width=50mm,height=50mm]{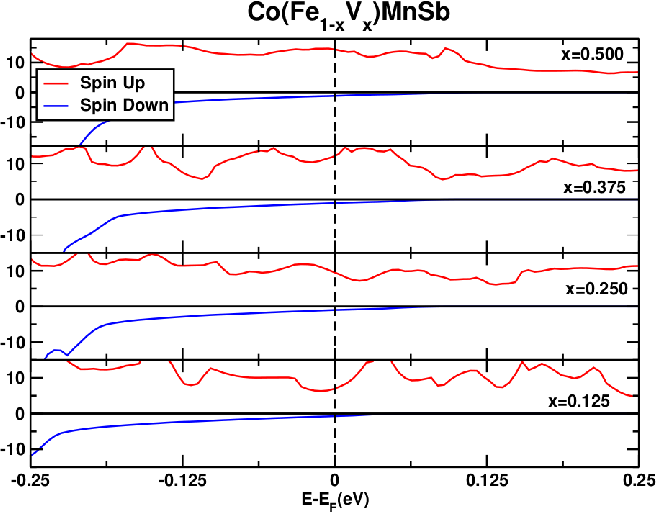}
\includegraphics[width=50mm,height=50mm]{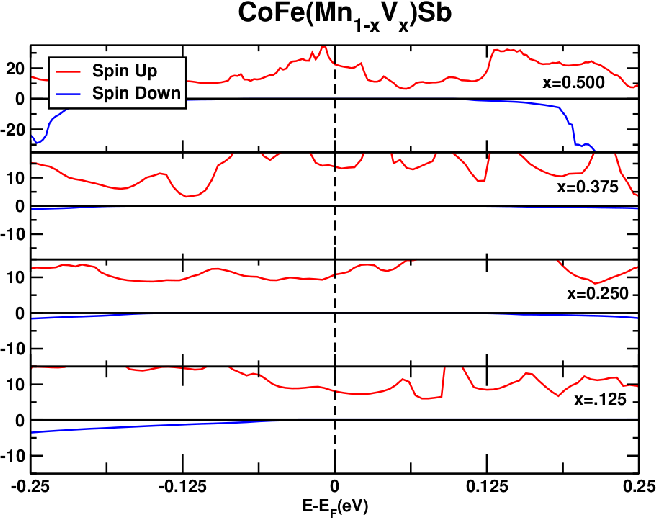}\\
\includegraphics[width=50mm,height=50mm]{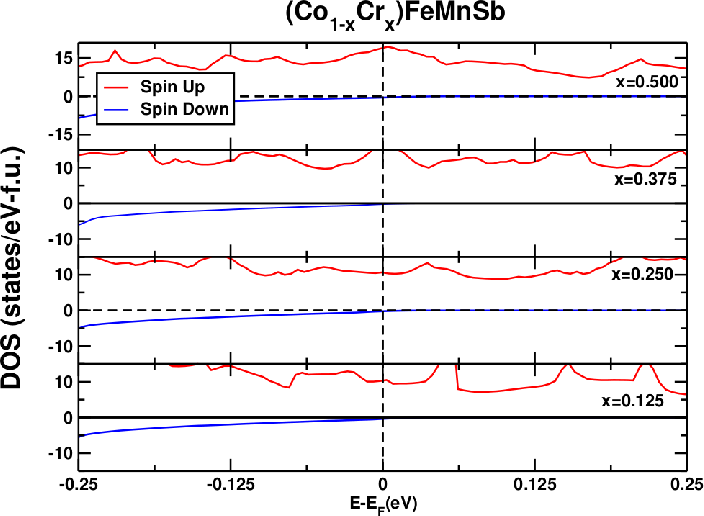}
\includegraphics[width=50mm,height=50mm]{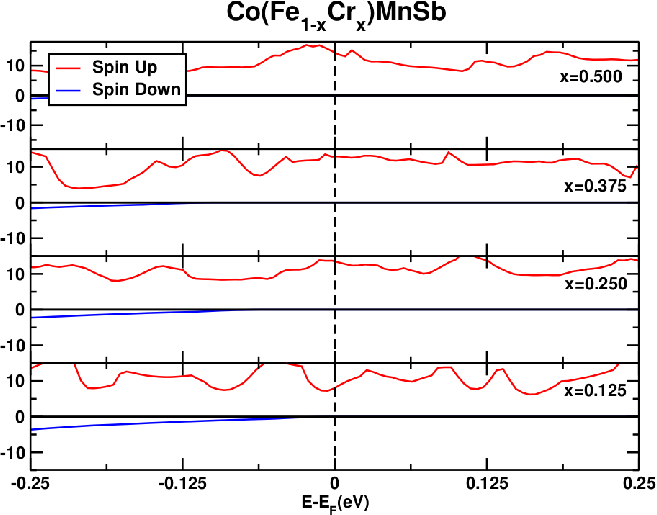}
\includegraphics[width=50mm,height=50mm]{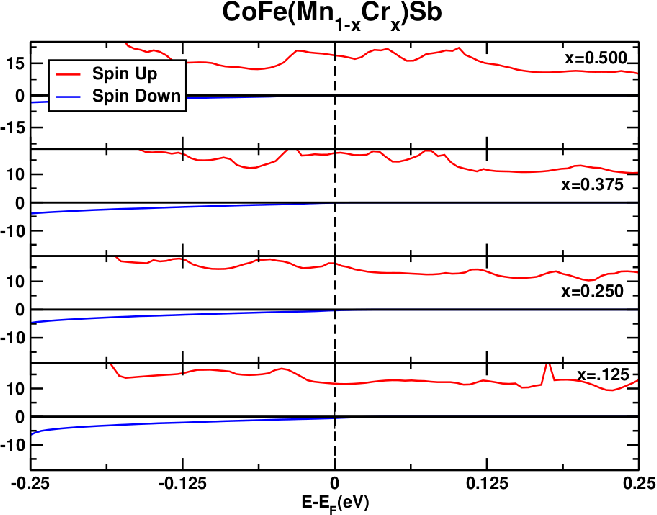}
\end{tabular}
\end{figure}
\begin{figure}
\centering
\begin{tabular}{@{}cccc@{}}
\includegraphics[width=50mm,height=50mm]{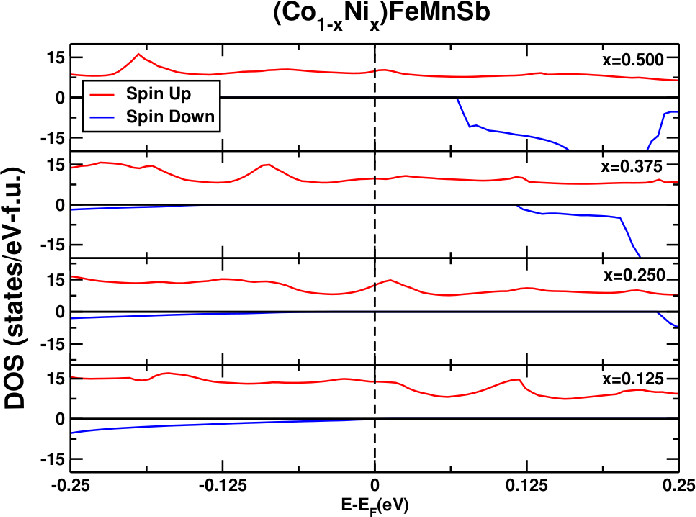}
\includegraphics[width=50mm,height=50mm]{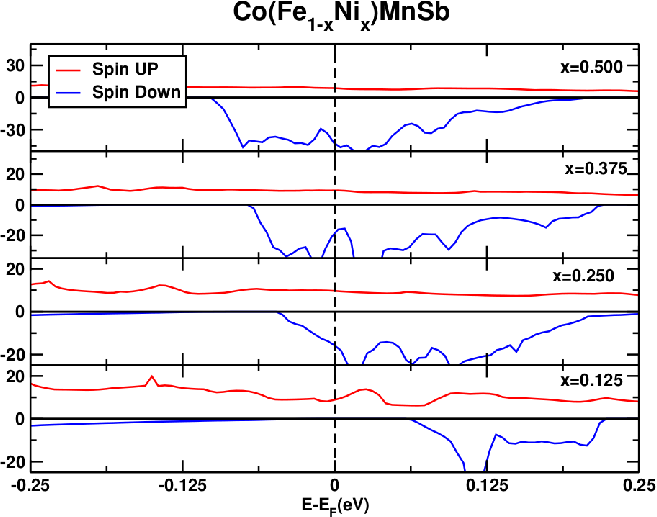}
\includegraphics[width=50mm,height=50mm]{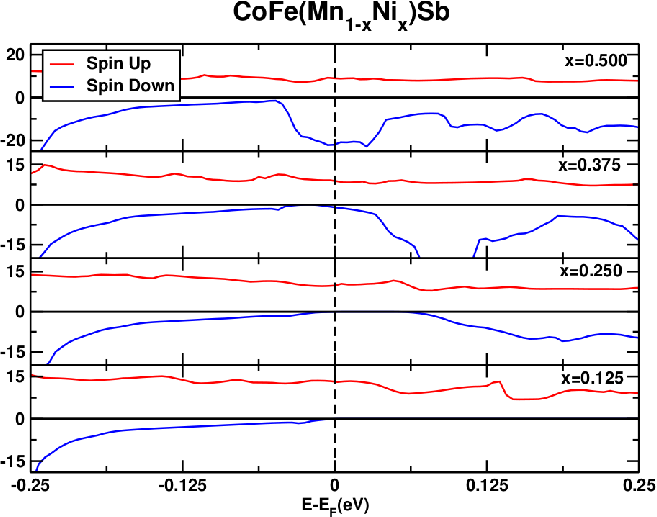}\\
\includegraphics[width=50mm,height=50mm]{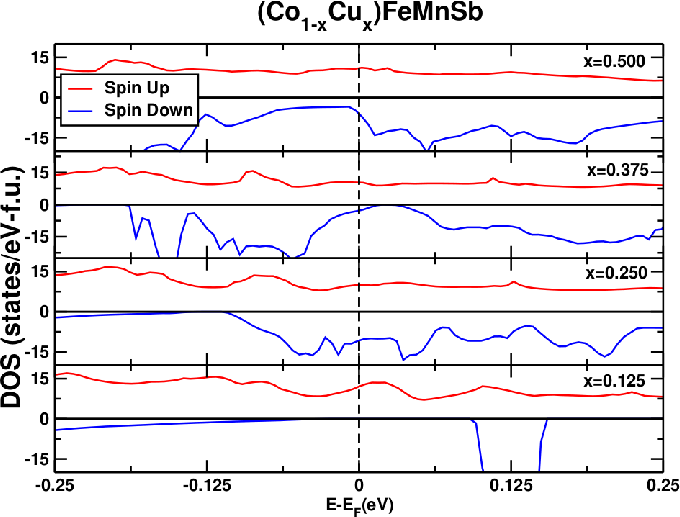}
\includegraphics[width=50mm,height=50mm]{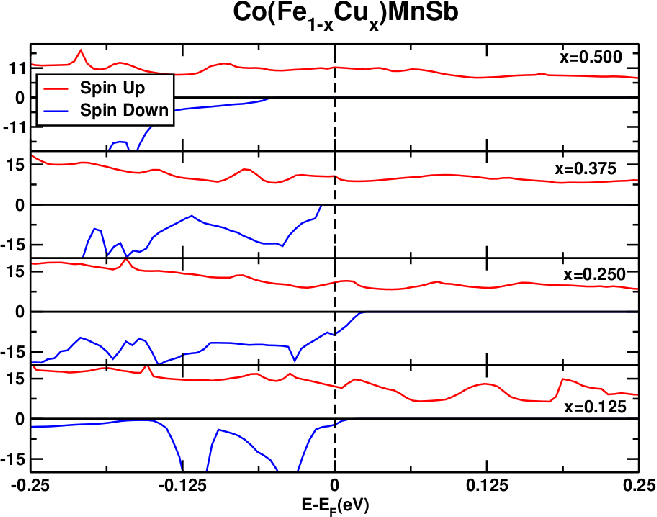}
\includegraphics[width=50mm,height=50mm]{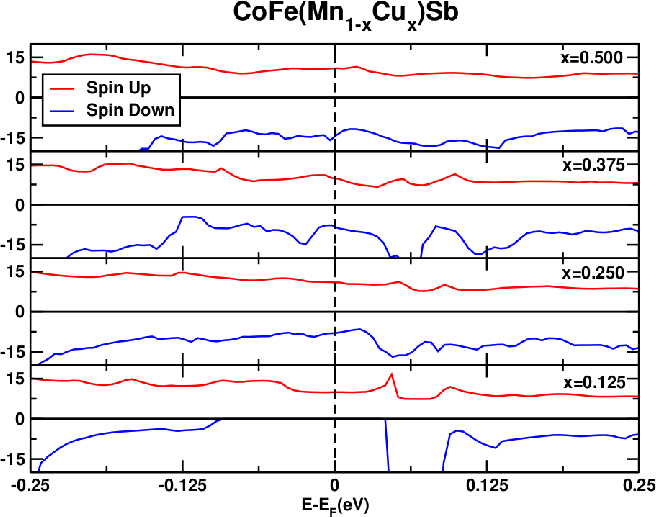}\\
\includegraphics[width=50mm,height=50mm]{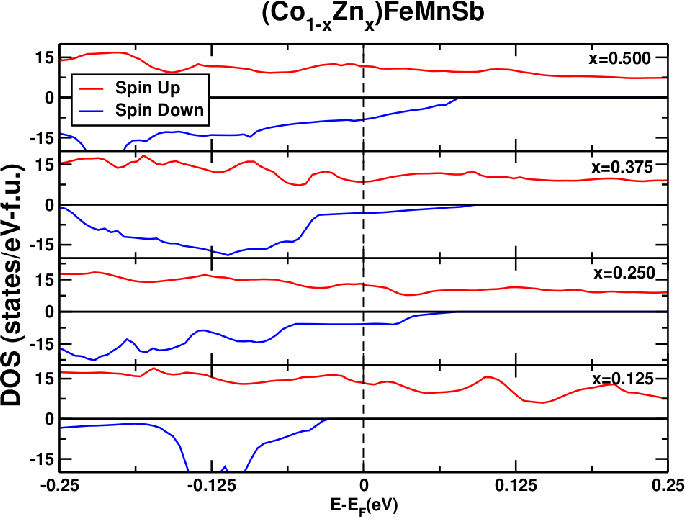}
\includegraphics[width=50mm,height=50mm]{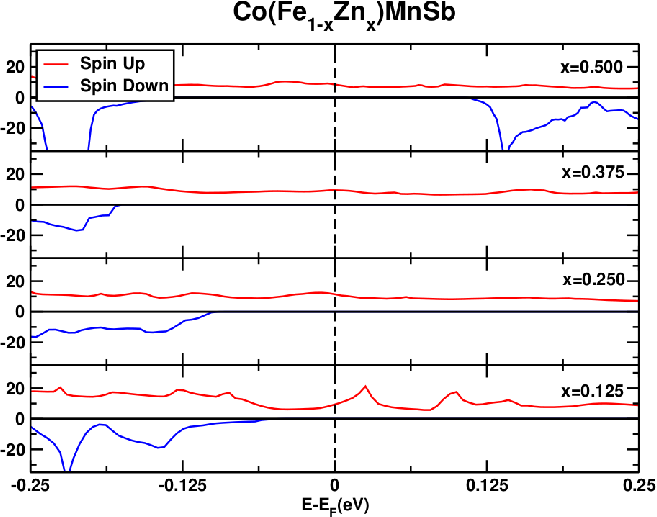}
\includegraphics[width=50mm,height=50mm]{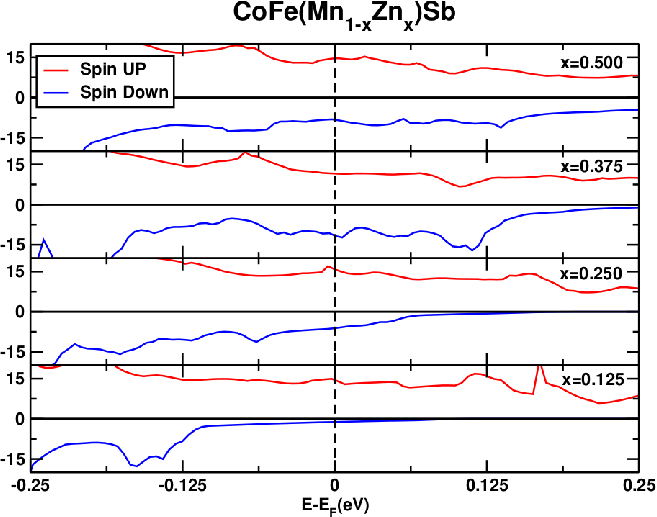}
\end{tabular}
\caption{Doping effect on electronic structure of CoFeMnSb at varies transitional metal sites. In each row left, middle and right plots show TM-alloyed at Co, Fe and Mn sites in CoFeMnSb respectively.}
\label{alloyant_dos}
\end{figure}

\clearpage
\vspace{-1cm}
\begin{figure}
\centering
\subfigure[]{\includegraphics[width=45mm,height=45mm]{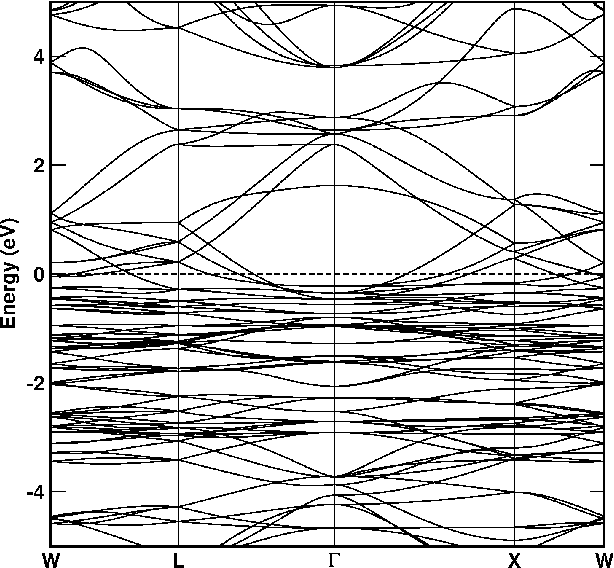}}
\subfigure[]{\includegraphics[width=45mm,height=45mm]{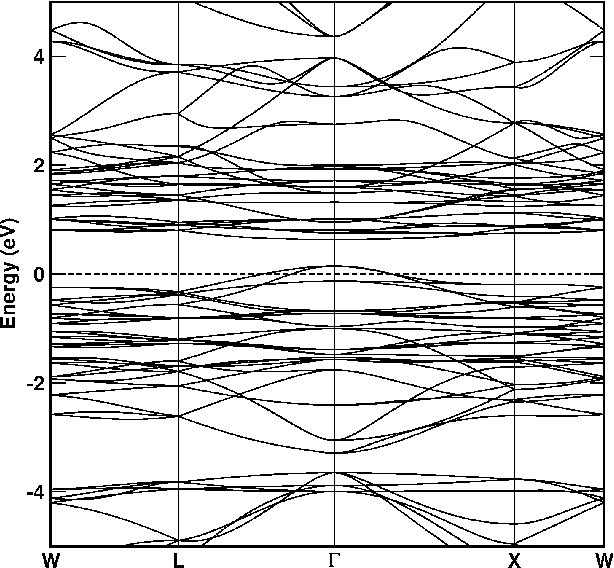}}\\
\subfigure[]{\includegraphics[width=45mm,height=45mm]{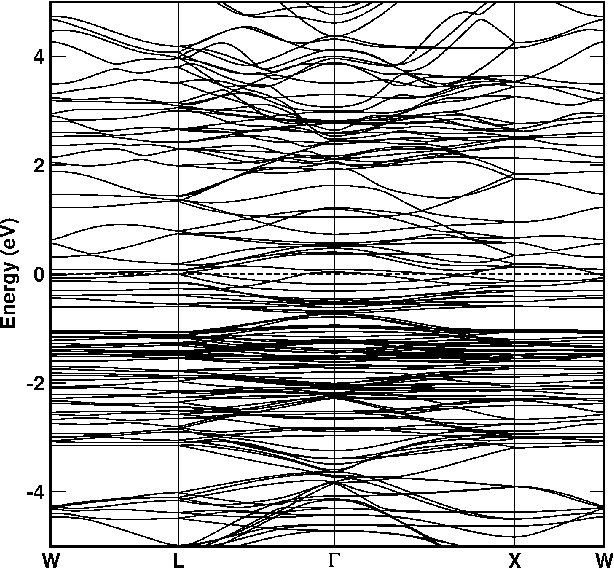}}
\subfigure[]{\includegraphics[width=45mm,height=45mm]{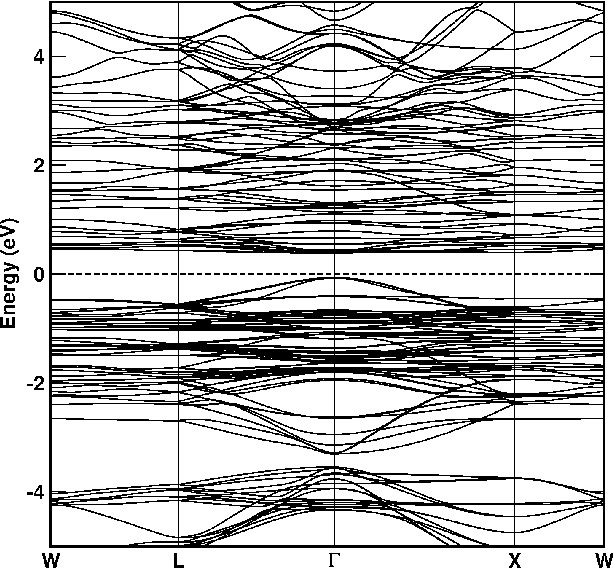}}\\
\subfigure[]{\includegraphics[width=45mm,height=45mm]{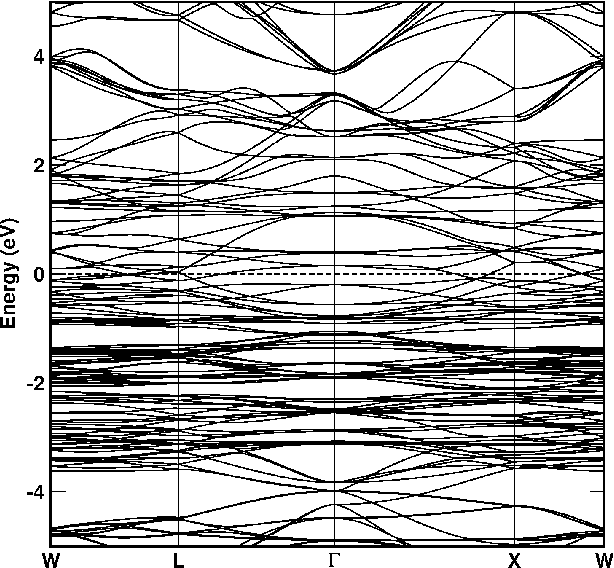}}
\subfigure[]{\includegraphics[width=45mm,height=45mm]{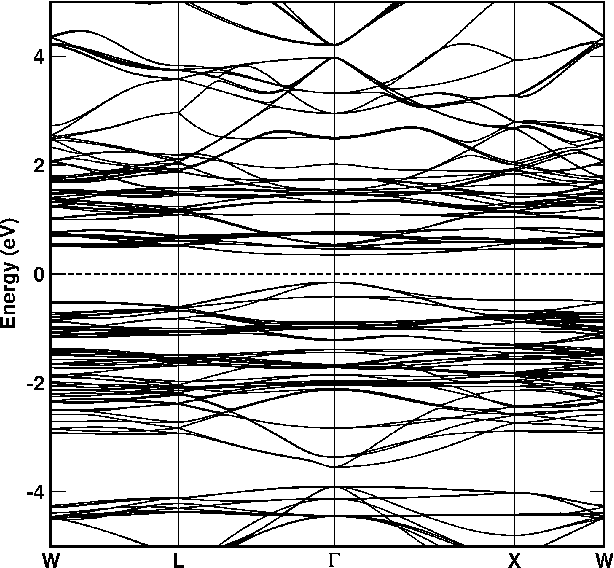}}\\
\subfigure[]{\includegraphics[width=45mm,height=45mm]{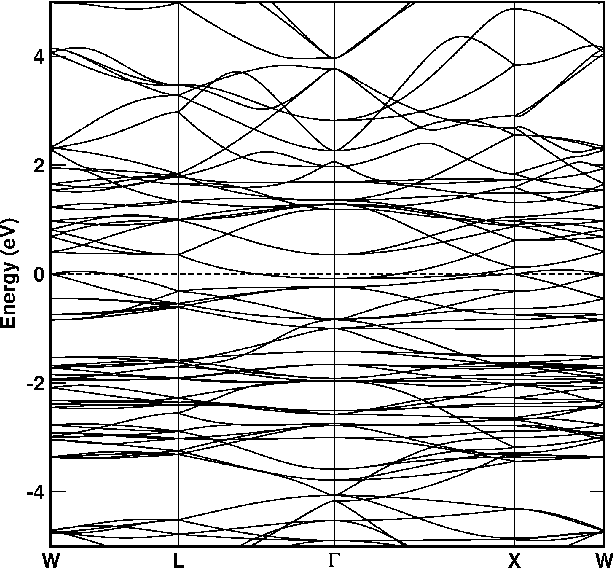}}
\subfigure[]{\includegraphics[width=45mm,height=45mm]{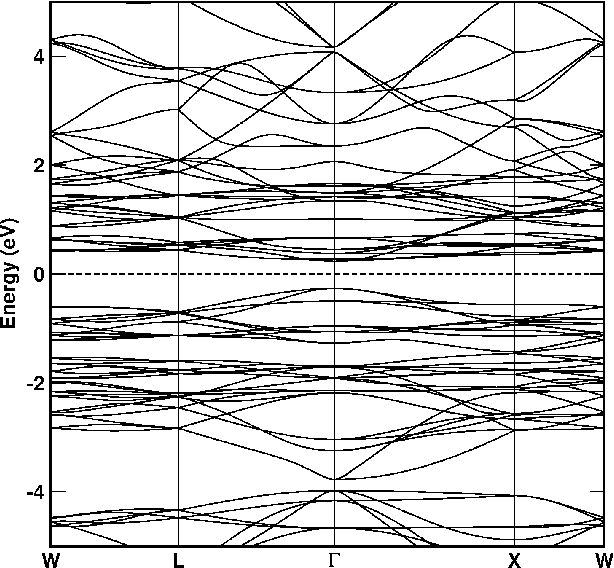}}\\
\caption{Band structure of bulk (a) spin up (b) spin dn, CoFe(Mn$_{0.25}$Sc$_{0.75}$)Sb (c) spin up (d) spin dn, CoFe(Mn$_{0.50}$Ti$_{0.50}$)Sb (e) spin up (f) spin dn, CoFeTiSb (g) spin up (h) spin dn.  }
\label{band}
\end{figure}

\end{document}